\documentclass[a4paper,11pt]{article}
\pdfoutput=1 

\usepackage{jheppub} 
\usepackage[T1]{fontenc} 
\usepackage{subcaption}
\usepackage{mathtools}
\graphicspath{ {graphs/} }

\preprint{YITP-SB-2017-56}

\title{\boldmath \LARGE On Mellin Amplitudes in SCFTs with Eight Supercharges}

\author[]{Xinan Zhou}

\affiliation[]{C. N. Yang Institute for Theoretical Physics, \\Stony Brook University, Stony Brook, 11794, NY, USA}

\emailAdd{xinan.zhou@stonybrook.edu}

\keywords{AdS/CFT correspondence, holographic four-point functions, Mellin amplitude, matter coupled gauged supergravity, 5d $F(4)$, Seiberg theories,  6d (1,0), E-string theory.}

\abstract{We extend the Mellin space techniques of \cite{Zhou:2017zaw} for computing holographic four-point correlation functions in maximally superconformal theories to theories with only eight Poincar\'e supercharges. The one-half BPS operators in these correlators are taken to be the superconformal primary in the $\mathcal{D}[k]$ multiplet (with $k=2$ corresponding to the flavor current multiplet), and transform in the adjoint representation of a flavor group $G$. Because of the smaller R-symmetry group $SU(2)$, each individual superconformal Ward identity  is less powerful. On the other hand, the constraining power is compensated in number by the different flavor channels in the four-point function. As  concrete test cases, we study the Seiberg theories in five dimensions and E-string theory in six dimensions at the large $N$ limit. We show that the flavor current multiplet four-point functions are fixed by superconformal symmetry up to two free parameters, which are proportional to the squared OPE coefficients for the flavor current multiplet and the stress tensor multiplet.

 }

\begin{document}
\maketitle
\flushbottom
\section{Introduction}
The AdS/CFT correspondence equates strongly coupled conformal field theories to gravitational theories in a higher dimensional spacetime. A wealth of information of the boundary CFT can be recovered by analyzing the four-point correlation functions which are dual to four-particle on-shell scattering amplitudes in the bulk AdS space. The computation of such bulk scattering amplitudes however is far from trivial. The standard algorithm to evaluate holographic correlators is straightforward but very cumbersome. To the leading non-trivial order in the large $N$ expansion,  one is instructed to compute a sum of tree-level Witten diagrams.  The requisite vertices can be read off from the effective action of the bulk theory, but Kaluza-Klein supergravity is devilishly complicated.\footnote{To appreciate its extraordinary difficulty, let us point out that the most general quartic vertices (which appear in the contact scattering diagrams) has only been worked out for IIB supergravity on $AdS_5\times S^5$ and the results took 15 pages \cite{Arutyunov:1999fb}.}  Though in certain backgrounds, there are streamlined procedures to evaluate exchange Witten diagrams as truncated sums of contact Witten diagrams \cite{DHoker:1999aa}, the evaluation  in position space is still very unwieldy. Moreover, the exchange diagrams proliferate very rapidly as the weights of the external operators are increased. Because of these difficulties, only a handful of four-point correlators have been computed in the past two decades using this recipe, to wit: a small number of examples in the $AdS_5\times S^5$ background \cite{Freedman:1998bj,Arutyunov:2000py,Arutyunov:2002fh,Arutyunov:2003ae,Berdichevsky:2007xd,Uruchurtu:2008kp,Uruchurtu:2011wh}; only the four-point of the stress tensor multiplet in $AdS_7\times S^4$ \cite{Arutyunov:2002ff}; and no results whatsoever for other backgrounds.\footnote{Except for certain very special cases which one can compute by other means: the extremal and next-to-extremal correlators in 4d $\mathcal{N}=4$ are given by their free field value thanks to non-renormalization theorems \cite{DHoker:1999jke,Bianchi:1999ie,Eden:1999kw,Erdmenger:1999pz,Eden:2000gg}; correlators in 3d, 4d and 6d with special R-symmetry polarization can be computed from chiral algebra (or 1d topological quantum mechanics in the 3d case) \cite{Beem:2013sza,Beem:2014kka,Chester:2014mea,Beem:2016cbd}.}

Only until recently did this sorry state of affairs begin to change, with a lot of progress in developing efficient calculational methods inspired by the bootstrap program.  A first approach,   referred to as the ``position space method'' in \cite{Rastelli:2016nze,Rastelli:2017udc}, was introduced to circumvent some of the difficulties in the traditional method. In this method one starts with an ansatz for the correlator which is a linear combination of all the possible Witten diagrams. By leveraging the special feature that in certain backgrounds (such as $AdS_5\times S^5$) all exchange Witten diagrams can be traded for finite sums of contact diagrams ($D$-functions), the ansatz reduces to a sum of finitely many $D$-functions. One then fixes all the relative coefficients by imposing the superconformal Ward identity, with no need of knowing the details of the effective supergravity action. Though much simpler than the conventional approach, this method also encounters unmanageable  computational difficulty when the KK levels of the external operators are increased. To have better analytic control, the idea of ``bootstrapping'' correlators using superconformal symmetry was then adapted to incorporate the Mellin representation formalism \cite{Mack:2009mi,Penedones:2010ue} which manifests the scattering nature of holographic correlators. In this second approach, one translates the task of computing four-point functions into solving an algebraic bootstrap problem for the Mellin amplitude \cite{Rastelli:2016nze,Rastelli:2017udc}. Superconformal symmetry is implemented by rewriting the position space {\it solution} of the superconformal Ward identity in Mellin space. The bootstrap problem is then formulated by combining superconformal constraints with additional self-consistency conditions of the Mellin amplitude.  By solving the algebraic problem, one obtains the most general one-half BPS four-point function with arbitrary Kaluza-Klein modes \cite{Rastelli:2016nze}. The same strategy also applies to $(2,0)$ theories in six dimensions and a similar problem was set up for eleven dimensional supergravity on $AdS_7\times S^4$  in \cite{Rastelli:2017ymc}. Unfortunately the generalization of this tactic to SCFTs in odd spacetime dimensions is difficult because the position space solution of the superconformal Ward identity does not have clear interpretation in Mellin space.  To this end, a complementary method was developed in \cite{Zhou:2017zaw} which translates only the superconformal Ward {\it identity}  into Mellin space. This approach presents a universal framework for imposing superconformal constraints on Mellin amplitudes in arbitrary dimensions. Using this method, the first full four-point for eleven dimensional supergravity on $AdS_4\times S^7$ was computed in \cite{Zhou:2017zaw}.\footnote{Recently this Mellin technique,  together with the exact CFT data from the supersymmetric localization, has been used to derive the leading M-theory correction beyond the supergravity limit  \cite{Chester:2018aca}.}
 
 All the prior work above has focused on correlators in backgrounds admitting maximal supersymmetry. Full correlators in non-maximal superconformal models, however, have never been studied holographically. The purpose of our paper is to explore the holographic computation of one-half BPS four-point functions with less supersymmetry, by extending the Mellin space techniques of \cite{Zhou:2017zaw}. More precisely we will set up the formalism for theories with only eight Poincar\'e supercharges.\footnote{This includes $\mathcal{N}=4$ in three dimensions, $\mathcal{N}=2$ in four dimensions, $F(4)$ in five dimensions and $\mathcal{N}=(1,0)$ in six dimensions. Two dimensions is special and is not discussed in this paper. Note that 5d $F(4)$ is non-maximal in the sense that it can be viewed as the conformal version of $5d$ $\mathcal{N}=1$ supersymmetry. The maximal amount of supersymmetry is $\mathcal{N}=2$ but it cannot be made conformal.} All these theories have at least $SU(2)$ R-symmetry -- the minimal R-symmetry required by the formalism of \cite{Dolan:2004mu} to derive superconformal Ward identities. Our motivation is two-fold. First, unlike the maximally supersymmetric theories which are conjectured to be unique, there is a zoo of superconformal theories with eight Poincar\'e supercharges. Among them, there are interesting isolated SCFTs in higher dimensions -- such as Seiberg theories in five dimensions \cite{Seiberg:1996bd} and E-string theory in six dimensions \cite{Ganor:1996mu,Seiberg:1996vs} -- for which holography is one of the few available tools to study them. Second, it has been demonstrated in \cite{Rastelli:2017ymc,Rastelli:2016nze,Rastelli:2017udc,Zhou:2017zaw} that maximal superconformal symmetry completely fixes the four-point correlators up to an overall constant. It is  curious to see to what extent non-maximal superconformal symmetry can constrain holographic correlators. In particular, if not all parameters in the amplitude ansatz are fixed by symmetry, what corresponds to the remaining ``degrees of freedom''? 
 
In this paper we will focus on the simplest four-point functions of the moment map operator, which is a one-half BPS operator sitting at the bottom of the flavor current multiplet $\mathcal{D}[2]$.\footnote{Here $\mathcal{D}[k]$ denotes the $\mathcal{D}$-type multiplets and $k$ is a positive integer. The bottom component of the $\mathcal{D}[k]$ multiplet is an one-half BPS scalar operator. It has conformal dimension  $\Delta=k\epsilon$ with $\epsilon=\frac{d-2}{2}$, and transforms in the $k$-fold symmetric traceless representation of the $SO(3) \simeq SU(2)_R$. See, {\it e.g.}, \cite{Buican:2016hpb,Cordova:2016emh} for a systematic analysis of unitary superconformal multiplets in various dimensions.} The same techniques also apply to the more general correlators of the $\mathcal{D}[k_i]$ multiplets, but the moment map four-point functions are already non-trivial. Such four-point functions, for example, are the objects considered in the superconformal bootstrap with eight Poincar\'e supercharges in four dimensions \cite{Beem:2014zpa}, five dimensions \cite{Chang:2017xmr} and six dimensions \cite{Chang:2017cdx}. We consider their holographic computation in two classes of isolated superconformal theories. The first family of theories  are the Seiberg theories in five dimensions which comes from $USp(2N)$ gauge theory coupled to $N_f\leq 7$ hypermultiplets in the fundamental representation and a single hypermultiplet in the antisymmetric representation. The ultraviolet fixed points have enhanced $E_{N_f+1}$ flavor symmetry and are referred to as the Seiberg exceptional superconformal field theories. The second theory is the E-string theory which arises as the infrared limit of $N$ M5 branes lying inside an ``end-of-the-world'' M9 brane, and has an $E_8$ flavor symmetry. Both theories have appropriate limits in which they can be described by supergravity models. The moment map four-point functions can be  computed from a  further consistent truncation of the theories as matter coupled gauged supergravity on $AdS_6$ and $AdS_7$. 

We use a bootstrap-like approach to compute them in Mellin space: we first formulate an ansatz for the Mellin amplitude which is consistent with the qualitative features of the supergravity theory; we then use the superconformal Ward identity to fix the parameters in the ansatz.  Compared to one-half BPS four-point functions in maximally superconformal theories, the computation of moment map four-point functions exhibits two new features. First of all, the four-point function has only one R-symmetry cross ratio because the R-symmetry is only $SU(2)$, which makes a single superconformal Ward identity weaker (see Section \ref{scfkin}). Secondly, because the moment map operators are in the adjoint representation of the flavor group, there are several flavor channels in the four-point function. Each flavor channel supplies a superconformal Ward identity so the overall constraining power is compensated by the number of identities. Remarkably, we find that all the parameters in the ansatz (in particular those parameterizing the quartic contact interactions) can be reduced to just two unknowns. These two remaining parameters have clear physical meaning: they are proportional to the squared OPE coefficients for the exchange of the flavor current multiplet and the stress tensor multiplet in the four-point function.  Alternatively, they can be interpreted to be proportional to the gauge coupling and gravitational coupling in the bulk.

One use of our results is to make contact with the aforementioned superconformal bootstrap program. Evidence was found that the Seiberg theories and E-string theories sit at the boundary of the space of unitary solutions of crossing. If such conjectures are true, then the bootstrap can potentially solve these theories using the extremal functional method \cite{ElShowk:2012hu,El-Showk:2014dwa}, and provide numerical estimates for the CFT data. Our result can provide a check for these bootstrap proposals at large central charge. From our Mellin amplitudes one can systematically extract anomalous dimensions and correction to the OPE coefficients for the low-lying operators. The comparison with future bootstrap results would be very interesting.\footnote{One such detailed analysis was recently performed in \cite{Chester:2018lbz} for the ABJM theory. The theory additionally contains one dimensional topological sectors \cite{Chester:2014mea,Beem:2016cbd} that can be accessed by supersymmetric localizations \cite{Dedushenko:2016jxl,Mezei:2017kmw,Agmon:2017lga}. OPE data was extracted from the Mellin amplitude in \cite{Zhou:2017zaw}, and showed remarkable agreement with the localization results \cite{Agmon:2017xes} and the $\mathcal{N}=8$ numeric bootstrap \cite{Chester:2014fya,Agmon:2017xes}. }

Although the paper mainly discusses the four-point functions of moment map operators, there are infinitely many more four-point functions to be considered with operators from the $\mathcal{D}[k_i]$ multiplets. For $k_i>2$, these one-half BPS operators are dual to the massive Kaluza-Klein modes of the bulk scalar field. It is a natural next step to use the formalism and techniques here to study these massive correlators. Studying these correlators will take us away from the regime of matter coupled gauged supergravity and start to explore the consequence of the internal manifold.  It would be interesting to obtain all the admissible four-point functions consistent with the symmetries,  and to identify the bulk meaning of the unfixed parameters. In this way, we are using the four-point functions to probe the ``landscape'' of  supergravity models in the bulk.
 
The rest of the paper is organized as follows. In Section \ref{scfkin} we review some basic superconformal kinematics of the moment map four-point function. We set up the Mellin formalism in Section \ref{Mellinrepandscfwi} and discuss how to implement superconformal constraint in Mellin space. Combining these ingredients, we discuss the strategy of computing four-point functions from holography in Section \ref{strategy}. We demonstrate the method in Seiberg theories for a number of flavor groups in Section \ref{selectedG} and in Section \ref{Estring} we apply the strategy to E-string theory in six dimensions.

\section{Some Superconformal Kinematics}\label{scfkin}
Superconformal symmetry organizes operators into superconformal multiplets. In particular, the flavor conserved current of a $d$-dimensional SCFT with eight Poincar\'e supercharges resides in the superconformal multiplet $\mathcal{D}[2]$ whose superconformal primary is the moment map operator $\mathcal{O}_{\alpha_1\alpha_2}^a$.  The moment map operator is a one-half BPS operator with conformal dimension $\Delta=d-2$.  It has R-charge $j_R=1$ under the $SU(2)_R$ R-symmetry (captured by $\alpha_1,\alpha_2=1,2$) and transforms in the adjoint representation under the flavor group $G$\footnote{We mostly focus on the non-mesonic symmetries, {\it e.g.}, $E_{N_f+1}$ for the Seiberg theories and $E_8$ for the E-string theory which have intrinsic gauge symmetry origin. The mesonic symmetry ($SU(2)$ for the Seiberg theories and the E-string theory) comes from the KK reduction of an internal manifold. The moment map operators of the $E_{N_f+1}$ and $E_8$ symmetry are singlets under the mesonic flavor symmetry.}(captured by the flavor index $a$). Holographically, the moment map operator is dual to a scalar field with $m^2=\Delta(\Delta-d)$ and the flavor conserved current is dual to a $G$ gauge field.  There are more one-half BPS
operators in $\mathcal{D}[k]$ multiplets. These operators $\mathcal{O}_{\alpha_1\ldots\alpha_k}^a$ are the scalar superconformal primaries and have conformal dimension $\Delta=k\epsilon$. They transform in the rank-$k$ symmetric tensor of the fundamental representation of $SU(2)_R$. We will also choose these operators (which we take as external operators) to be in the adjoint representation of the flavor group as for the $k=2$ operator.\footnote{Here we suppressed the mesonic flavor symmetries under which the $\mathcal{D}[k]$ multiplets with $k>2$ transform non-trivially. Their symmetry indices can be treated in the same way as for the $SU(2)_R$ R-symmetry indices so we have left them out to avoid overloading the notation. Of course, under the remaining flavor group there are $\mathcal{D}[k]$ multiplets in other representation, {\it e.g.}, as multi-trace operators of $\mathcal{D}[k]$, but they do not correspond to supergravity fields.} Such operators are dual in the bulk to the massive modes of a KK tower whose lowest component is $\mathcal{O}_{\alpha_1\alpha_2}^a$.

We are interested in the four-point functions of such one-half BPS operators \begin{equation}
\langle \mathcal{O}_{\alpha_1\ldots\alpha_k}^a(x_1)\mathcal{O}_{\beta_1\ldots\beta_k}^b(x_2)\mathcal{O}_{\gamma_1\ldots\gamma_k}^c(x_3)\mathcal{O}_{\delta_1\ldots\delta_k}^d(x_4)\rangle\;.
\end{equation}
For simplicity we have taken all the external weights to be equal to $k$, although the generalization to unequal weights is straightforward. By superconformal symmetry, this four-point function uniquely fixes all the four-point correlators of the superconformal descendants in the $\mathcal{D}[k]$ multiplet.\footnote{This follows from a simple counting. The superfield of a one-half BPS multiplet depends on four fermionic coordinates. The ``super correlator'' of four such superfields then contains sixteen fermionic coordinates,  which equals the number of fermionic generators in the superconformal algebra. One can therefore use the superconformal symmetry to restore all the fermionic coordinates in the super correlator from the ``bottom correlator''. However, ``harmonic singularities'' may arise from this procedure. Requiring the absence of these singularities gives the superconformal Ward identity (\ref{scfwiposi}) for the four-point correlator \cite{Dolan:2004mu}.} Because $\mathcal{O}_{\alpha_1\ldots\alpha_k}^a$ transforms in the rank-$k$ symmetric traceless representation of $SU(2)_R$, we can conveniently keep track of the R-symmetry by contracting them with a (commuting) two-component spinor $t^\alpha$. The four-point function then becomes a function of both the spacetime coordinates $x_i$ and the R-symmetry coordinates $t_i$
\begin{equation}
G^{abcd}(x_i,t_i)\equiv\langle \mathcal{O}^a(x_1,t_1)\mathcal{O}^b(x_2,t_2)\mathcal{O}^c(x_3,t_3)\mathcal{O}^d(x_4,t_4) \rangle
\end{equation}
where $\mathcal{O}^a(x,t)\equiv \mathcal{O}^a_{\alpha_1\ldots\alpha_k}(x)t^{\alpha_1}\ldots t^{\alpha_k}$. Using the covariance under the conformal group and R-symmetry group, we can further write the four-point function as a function only of the cross ratios
\begin{equation}\label{calG}
G^{abcd}(x_i,t_i)\equiv \left(\frac{t_{12}t_{34}}{x_{12}^{2\epsilon} x_{34}^{2\epsilon} }\right)^k\mathcal{G}^{abcd}(U,V;\alpha)\;,
\end{equation}
with $x_{ij}\equiv x_i-x_j$, $t_{ij}\equiv t_1^\alpha t_2^\beta\epsilon_{\alpha\beta}$ and $\epsilon=\frac{d-2}{2}$. Here we defined the usual conformal cross ratios
\begin{equation}
U=\frac{x_{12}^2x_{34}^2}{x_{13}^2x_{24}^2}\;,\quad\quad V=\frac{x_{14}^2x_{23}^2}{x_{13}^2x_{24}^2}\;,
\end{equation}
and similarly the R-symmetry cross ratio
\begin{equation}
\alpha\equiv \frac{t_{13}t_{24}}{t_{12}t_{34}}\;.
\end{equation}
Notice due to the identity
\begin{equation}\label{tijiden}
t_{12}t_{34}-t_{13}t_{24}+t_{14}t_{23}=0\;,
\end{equation}
there is only one independent R-symmetry cross ratio.\footnote{For R-symmetry group that is locally isomorphic to $SO(n)$ with $n>3$, there is another independent R-symmetry cross ratio $\beta=\frac{t_{14}t_{23}}{t_{12}t_{34}}$. In the $SO(3)$ case, these two cross ratios become linearly dependent $\beta-\alpha=1$ because of the identity (\ref{tijiden}).}
 It is also not hard to see that $\mathcal{G}^{abcd}(U,V;\alpha)$ is a degree-$k$ polynomial in $\alpha$. 

So far in (\ref{calG}) we have only exploited the bosonic symmetries of the superconformal group. The fermionic generators further yields the following additional constraints in the form of {\it superconformal Ward identity} \cite{Dolan:2004mu}\footnote{For superconformal groups with R-symmetry $SO(n)$ and $n>3$, the superconformal Ward identity reads
\begin{equation}
(\chi\partial_\chi-\epsilon a\partial_a)\mathcal{G}(\chi,\chi';a,a')\big|_{a=1/\chi}=0
\end{equation}
where $\alpha=a a'$ and $\beta=(1-a)(1-a')$. The untwisted R-symmetry cross ratio $a$ is a spectator in the above identity and the left hand side is a degree-$k$ polynomial of $a$. The superconformal Ward identity holds for  the coefficient of each individual power of $a$ and therefore gives $k+1$ identities.}
\begin{equation}\label{scfwiposi}
\left(\chi\partial_\chi-\epsilon\alpha\partial_\alpha\right)\mathcal{G}^{abcd}(\chi,\chi';\alpha)\big|_{\alpha=1/\chi}=0\;.
\end{equation}
In the above, we have made a convenient change of variables 
\begin{equation}
U=\chi\chi'\;,\quad\quad V=(1-\chi)(1-\chi')\;.
\end{equation}

Another important property of the four-point function is its transformation under crossing. To write down its crossing equations, it is  convenient to project the four-point function $\mathcal{G}^{abcd}$ into the flavor irreducible representations that appear in the s-channel tensor product $\mathrm{Adj}_1\otimes \mathrm{Adj}_2$ with the projection matrix $P_I^{ab|cd}$
\begin{equation}
\mathcal{G}^{abcd}(U,V;\alpha)=\sum_{\mathcal{R}_I\in \mathbf{adj}\otimes \mathbf{adj}}P_I^{ab|cd}\mathcal{G}_I(U,V;\alpha)\;.
\end{equation}
Each flavor channel $\mathcal{R}_I$ independently obeys the superconformal Ward identity (\ref{scfwiposi}), therefore
\begin{equation}\label{scfwiposiproj}
\left(\chi\partial_\chi-\epsilon\alpha\partial_\alpha\right)\mathcal{G}_{I}(\chi,\chi';\alpha)\big|_{\alpha=1/\chi}=0\;.
\end{equation}
The matrix $P_I^{ab|cd}$ satisfies the following properties
\begin{eqnarray}\label{Pprop}
\nonumber P_I^{ab|cd}&=&(-1)^{|\mathcal{R}_I|}P_I^{ba|cd}\;,\\
\nonumber P_I^{ab|cd}&=&P_I^{cd|ab}\;,\\
P_I^{ab|cd}P_J^{ab|cd}&=&\delta_{IJ}\,\mathrm{dim}(\mathcal{R}_I)
\end{eqnarray}
where $|\mathcal{R}_I|=0$ if $\mathcal{R}_I$ is a symmetric representation and $|\mathcal{R}_I|=1$ if $\mathcal{R}_I$ is an anti-symmetric representation. Using the crossing invariance of (\ref{calG}), we can derive the following crossing equations.

\noindent\underline{s-channel to t-channel  (swapping 1 and 3):}

\begin{equation}\label{posistot}
\sum_J (F_t)_I{}^J\mathcal{G}_J(U,V;\alpha)=\left(\frac{U}{V}\right)^{k\epsilon}(\alpha-1)^k\mathcal{G}_I\left(V,U;\frac{\alpha}{\alpha-1}\right)\;,
\end{equation}

\noindent\underline{s-channel to u-channel (swapping 1 and 4):}

\begin{equation}\label{posistou}
\sum_J (F_u)_I{}^J\mathcal{G}_J(U,V;\alpha)=U^{k\epsilon}\alpha^k\mathcal{G}_I\left(\frac{1}{U},\frac{V}{U};\frac{1}{\alpha}\right)
\end{equation}
where we have defined the flavor crossing matrices
\begin{equation}
(F_t)_I{}^J\equiv\frac{1}{\mathrm{dim}(\mathcal{R}_I)}P_I^{cb|ad}P_J^{ab|cd}\;,\quad (F_u)_I{}^J\equiv\frac{1}{\mathrm{dim}(\mathcal{R}_I)}P_I^{db|ca}P_J^{ab|cd}\;.
\end{equation}
Using the basic properties (\ref{Pprop}) of the projection matrix, we find the two $F$-matrices are related by
\begin{equation}
(F_u)_I{}^J=(F_t)_I{}^J(-1)^{|\mathcal{R}_I|}(-1)^{|\mathcal{R}_J|}
\end{equation}
where no summations over $I$ and $J$ are performed. One can verify that
\begin{equation}\label{FF}
(F_t)^2=(F_u)^2=1\;.
\end{equation}

\section{Mellin Representation and Mellin Superconformal Ward Identity}\label{Mellinrepandscfwi}
A very useful language for studying conformal correlators, especially in the context of holography, is the Mellin representation formalism \cite{Mack:2009mi,Penedones:2010ue}.\footnote{See \cite{Penedones:2010ue,Fitzpatrick:2011ia,Paulos:2011ie,Costa:2012cb,Fitzpatrick:2011hu,Nandan:2011wc,Costa:2014kfa,Goncalves:2014rfa,Goncalves:2014ffa,Rastelli:2016nze,Alday:2017gde,Rastelli:2017ecj,Rastelli:2017udc,Faller:2017hyt,Rastelli:2017ymc,Zhou:2017zaw,Chen:2017xdz,Goncalves:2018fwx} for applications at tree level and \cite{Fitzpatrick:2011hu,Aharony:2016dwx,Yuan:2018qva} at loop level. Another interesting application of the Mellin representation formalism is in the bootstrap program, see \cite{Paulos:2016fap,Gopakumar:2016wkt,Gopakumar:2016cpb,Dey:2016mcs,Dey:2017oim,Dey:2017fab,Golden:2017fip}. }  We will refer the reader to Section 3 of \cite{Rastelli:2017udc} for a review of this formalism and the conventions used here.

Using this representation, we write the four-point function as a double inverse Mellin transformation
\begin{equation}\label{mellincalG}\small
\mathcal{G}^{abcd}(U,V;\alpha)=\int_{-i\infty}^{i\infty}\frac{ds}{4\pi i}\frac{dt}{4\pi i}U^{\frac{s}{2}}V^{\frac{t}{2}-k\epsilon}\mathcal{M}^{abcd}(s,t;\alpha)\Gamma^2\left[\frac{2k\epsilon-s}{2}\right]\Gamma^2\left[\frac{2k\epsilon-t}{2}\right]\Gamma^2\left[\frac{2k\epsilon-u}{2}\right]
\end{equation}
where $s$, $t$ and $u$ are the analogue of Mandelstam variables satisfying the condition $s+t+u=4k\epsilon$. $\mathcal{M}^{abcd}(s,t)$ is called the {\it Mellin amplitude} and carries the flavor indices $a$, $b$, $c$, $d$. It will also be useful to project the Mellin amplitude into the irreducible representations appearing in the tensor product of two adjoint representations
\begin{equation}
\mathcal{M}^{abcd}(s,t)=\sum_{\mathcal{R}_I\in \mathbf{adj}\otimes \mathbf{adj}}P_I^{ab|cd}\mathcal{M}_I(s,t)\;.
\end{equation}

The Mellin amplitude is crossing symmetric. We can derive its crossing equations by substituting the Mellin representation (\ref{mellincalG}) into (\ref{posistot}) and (\ref{posistou})
\begin{eqnarray}\label{mellincrossing}
\nonumber \mathcal{M}_I(s,t;\alpha)&=&(\alpha-1)^k\sum_J (F_t)_I{}^J\mathcal{M}_J\left(t,s;\frac{\alpha}{\alpha-1}\right)\\
&=& \alpha^k\sum_J (F_u)_I{}^J\mathcal{M}_J\left(u,t;\frac{1}{\alpha}\right)\;.
\end{eqnarray}
In deriving these relations we have also used (\ref{FF}).

Moreover, the Mellin amplitude of a superconformal field theory is constrained by intricate relations as a result of the underlying superconformal symmetry.  These relations take the from of difference identities which can be derived from the position space superconformal Ward identity (\ref{scfwiposiproj}), as was explained in \cite{Zhou:2017zaw}. Here we give a brief review of how to derive these relations. To be explicit, we specialize our discussion to $k=2$ which corresponds to the four-point functions of the flavor current multiplet.

For starters, we decompose the four-point function $\mathcal{G}_I(U,V;\alpha)$ into monomials of the R-symmetry cross ratio $\alpha$
\begin{equation}
\mathcal{G}_I(U,V;\alpha)=\mathcal{G}_I^0(U,V)+\alpha \mathcal{G}_I^1(U,V)+\alpha^2 \mathcal{G}_I^2(U,V)\;.
\end{equation} 
To apply the superconformal Ward identity (\ref{scfwiposiproj}), we write 
\begin{equation}\label{chidchi}
\chi\frac{\partial}{\partial \chi}=U\frac{\partial}{\partial U}+V\frac{\partial}{\partial V}-\frac{1}{1-\chi}V\frac{\partial}{\partial V}\;,
\end{equation}
but postpone the evaluation of $U\partial_U$ and $V\partial_V$. After multiplying with $(1-\chi)\chi^2$, the left side of (\ref{scfwiposiproj}) becomes a degree-3 polynomial of $\chi$
\begin{equation}\label{chiiden}
f_0(U,V)+\chi f_1(U,V)+\chi^2 f_2(U,V)+\chi^3 f_3(U,V)=0
\end{equation}
where
\begin{eqnarray}
\nonumber f_0&=& (-2\epsilon+U\partial_U)\mathcal{G}_I^2\;,\\
\nonumber f_1&=&\left(-\epsilon+U\partial_U\right)\mathcal{G}_I^1-(-2\epsilon+U\partial_U+V\partial_V)\mathcal{G}_I^2\;,\\
\nonumber f_2&=&U\partial_U\mathcal{G}_I^0-U\partial_U\mathcal{G}_I^1+\left(\epsilon-V\partial_V\right)\mathcal{G}_I^1\;,\\
f_3&=&-(U\partial_U+V\partial_V)\mathcal{G}_I^0\;.
\end{eqnarray}
Notice $U=\chi\chi'$, $V=(1-\chi)(1-\chi')$ are invariant under the exchange of $\chi$ and $\chi'$. This gives us another identity from (\ref{chiiden})
\begin{equation}\label{chiprimeiden}
f_0(U,V)+\chi' f_1(U,V)+\chi'^2 f_2(U,V)+\chi'^3 f_3(U,V)=0\;.
\end{equation}
We add up (\ref{chiiden}) and (\ref{chiprimeiden})
\begin{equation}\label{chichiprimeiden}
2f_0(U,V)+(\chi+\chi') f_1(U,V)+(\chi^2+\chi'^2) f_2(U,V)+(\chi^3+\chi'^3) f_3(U,V)=0\;.
\end{equation}
and notice all the $\chi^n+\chi'^n$ factors can be written as polynomials of $U$ and $V$
\begin{eqnarray}
\chi+\chi'&=&U-V+1\;,\quad \chi^2+\chi'^2=U^2+V^2-2UV-2V+1\;,\\
\chi^3+\chi'^3&=& U^3-V^3-3U^2V+3UV^2-3UV+3V^2-3V+1\;.
\end{eqnarray}
The fact these factors are polynomials of $U$ and $V$ becomes particularly convenient when we use the Mellin representation (\ref{mellincalG}) for each R-symmetry partial amplitude 
\begin{equation}\label{mellincalGIRmono}
\mathcal{G}_I^i(U,V)=\int_{\mathcal{C}}\frac{ds}{4\pi i}\frac{dt}{4\pi i}U^{\frac{s}{2}}V^{\frac{t}{2}-2\epsilon}\mathcal{M}^i_I(s,t)\Gamma^2\left[\frac{4\epsilon-s}{2}\right]\Gamma^2\left[\frac{4\epsilon-t}{2}\right]\Gamma^2\left[\frac{4\epsilon-u}{2}\right]
\end{equation}
where $i=0,1,2$. We notice the multiplicative monomial $U^mV^n$ can be absorbed into the integrand by shifting the $s$ and $t$ variables
\begin{equation}
s\to s-2m\;,\quad t\to t-2n\;.
\end{equation}
This promotes the monomial into an operator 
\begin{equation}\label{shiftop}
\underline{\widehat{U^mV^n}}:\; \mathcal{M}_I^i(s,t)\to {\mathcal{M}'}_I^i(s,t)
\end{equation}
where 
\begin{equation}
{\mathcal{M}'}_I^i(s,t)=\mathcal{M}_I^i(s-2m,t-2n)\left(\frac{4\epsilon-s}{2}\right)_m^2\left(\frac{4\epsilon-t}{2}\right)_n^2\left(\frac{4\epsilon-u}{2}\right)_{-m-n}^2\;.
\end{equation}
The Pochhammer symbols arise as a result of the same Gamma function factor that we have kept in the integrand
\begin{equation}\label{monomialops}
U^mV^n\mathcal{G}_I^i(U,V)=\int_{\mathcal{C}'} \frac{ds}{4\pi i}\frac{dt}{4\pi i}U^{\frac{s}{2}}V^{\frac{t}{2}-2\epsilon}{\mathcal{M}'}^i_I(s,t)\Gamma^2\left[\frac{4\epsilon-s}{2}\right]\Gamma^2\left[\frac{4\epsilon-t}{2}\right]\Gamma^2\left[\frac{4\epsilon-u}{2}\right]\;.
\end{equation}\label{shifted}
Moreover, it is easy to see that under the representation (\ref{mellincalGIRmono}) 
\begin{eqnarray}
\nonumber U\frac{\partial}{\partial U}&:&\; {\mathcal{M}}^i_I(s,t)\to {\mathcal{M}}^i_I(s,t)\times \frac{s}{2}\;,\\
V\frac{\partial}{\partial V}&:&\; {\mathcal{M}}^i_I(s,t)\to {\mathcal{M}}^i_I(s,t)\times \left(\frac{t}{2}-2\epsilon\right)\;.
\end{eqnarray}
All in all, (\ref{chichiprimeiden}) translates in the Mellin space into a difference identity relating different R-symmetry components of the Mellin amplitude.\footnote{One may worry about subtleties related to contours because they are shifted as well under the action of shift operators. However no issue arises in the cases we consider in this paper, as we we can see from the following simple argument. Schematically, (\ref{chichiprimeiden}) can be written as 
\begin{equation}
 F_1(U,V)+F_2(U,V)+\ldots+F_n(U,V)=0
\end{equation}
where each $F_i(U,V)$ has some inverse Mellin representation of which the fundamental domain is $\mathcal{D}_i$. We perform a double Mellin transformation on both sides with respect to $U$ and $V$, with the powers of $U$ and $V$ chosen to be inside $\mathcal{D}_1$. From $F_1$ we simply get the Mellin transform $\mathcal{M}[F_1]$. But from $F_i$ with $i\neq 1$, the integral should be understood as the analytic continuation of $\mathcal{M}[F_i]$ from $\mathcal{D}_i$ to $\mathcal{D}_1$. Since the analytic continuation in our case will be unique (no branch cuts), it just becomes a statement of the integrands
\begin{equation}
\nonumber \mathcal{M}[F_1]+\mathcal{M}[F_2]+\ldots+\mathcal{M}[F_n]=0\;.
\end{equation}
} The above discussion holds for any flavor channel, and we therefore have as many identities as the irreducible representations that appear in the tensor product of two adjoint representations of the flavor group.

\section{Flavor Current Four-Point Function of Seiberg Theories}

In this section we present an efficient method for computing holographic four-point functions for the flavor current multiplet in Mellin space. The method that we will discuss in this section is a bootstrap-like approach: after writing down a general ansatz which captures the essential qualitative features of the supergravity theory, we use the superconformal symmetry to solve the parameters. The discussion will be embedded into the context of a specific class of theories, namely, the Seiberg theories in five dimensions \cite{Seiberg:1996bd}. However, the strategy can easily be applied to other theories in different spacetime dimensions, {\it e.g.}, the E-string theory that we will discuss in Section \ref{Estring}, with obvious modifications.

Before we go on to discuss the method, let us briefly review the Seiberg theories and their holographic duals. The Seiberg theories are a class of five dimensional superconformal field theories argued to exist as the ultraviolet fixed point of supersymmetric gauge theories. The gauge theories have gauge group $USp(2N)$ and are coupled to $N_f\leq 7$ hypermultiplets in the fundamental representation and a single hypermultiplet in the antisymmetric representation. They also can be constructed in Type I' string theory by a D4-D8/O8-brane setup. In this setup, we place two orientifold O8-planes at $x^9=0$ and $x^9=\pi$. We let $0\leq N_f<8$ D8-branes coincide with the O8 at $x^9=0$ and $16-N_f$ coincide with the other. On top of that, we place $N$ D4-branes in the $\{x^0,x^1,x^2,x^3,x^4\}$ directions. The Seiberg theories have flavor symmetries $E_{N_f+1}\times SU(2)_M$ where the first factor $E_{N_f+1}$\footnote{$E_1=SU(2)$, $E_2=SU(2)\times U(1)$, $E_3=SU(3)\times SU(2)$, $E_4=SU(5)$, $E_5=Spin(10)$.} is enhanced from the $SO(2N_f)$ symmetry of the $N_f$ D8-branes and $U(1)_I$ instanton particle symmetry. The mesonic $SU(2)_M$ comes from rotations in $\{x^5,x^6,x^7,x^8\}$ directions. The decoupling limit suggests a duality between the Seiberg theories and type I' string theory on $\mathcal{M}_6\times_w HS^4$ where $\mathcal{M}_6$ is an asymptotically locally $AdS_6$ space \cite{Seiberg:1996bd,Ferrara:1998gv}. In an appropriate low energy regime, the gravity side is described by Type I' supergravity \cite{Polchinski:1995df}. In particular the region between two D8 branes is sufficiently described by massive Type IIA supergravity \cite{Romans:1985tz}. There is also a consistent truncation of the theory into Romans $F(4)$ gauged supergravity \cite{Romans:1985tw} in $AdS_6$ coupled to matter vector multiplets with gauge group $E_{N_f+1}$ \cite{DAuria:2000afl,Andrianopoli:2001rs}. This latter limit that is relevant to the holographic computation of the flavor current multiplet four-point function and will be mainly considered in the following. The field content of the matter coupled gauged supergravity is organized into two multiplets. The stress tensor multiplet is neutral under the $E_{N_f+1}$ flavor symmetry and contains the stress tensor, the $SU(2)_R$ current, a dimension-3 scalar and their fermionic siblings. The flavor current multiplet transforms in the adjoint representation of $E_{N_f+1}$ and contains the $E_{N_f+1}$ flavor currents, the moment map operator, a dimension-4 $SU(2)_R$ singlet scalar and their fermionic counterparts. Because the external operators are scalars, only scalar, vector and graviton fields can be exchanged in the Witten diagrams.

This rest of the section is organized as follows. In Section \ref{strategy} we outline the general strategy of the computation. We formulate the ansatz in Mellin space in  Section \ref{secansatz} and  in Section \ref{secsolward} we discuss how to use the superconformal Ward identities for the Mellin amplitude. In Section \ref{selectedG} we implement the method and  give concrete solutions to four-point functions for selected flavor groups. 

\subsection{The Strategy of Computation}\label{strategy}

\subsubsection{The Ansatz for the Mellin Amplitude}\label{secansatz}
Using the standard recipe of AdS/CFT, the flavor current four-point function of the boundary CFT is computed by Witten diagram expansion in the bulk. The leading contribution in $1/N$ are trivially known from generalized free fields. Holographically it is computed from the disconnected diagrams which factorizes into products of two-point functions. The connected correlator is dominated by tree diagrams which consist of exchange Witten diagrams and contact Witten diagrams. Accordingly, the Mellin amplitude of the connected correlator can be written as 
\begin{equation}\label{ansatz}
\mathcal{M}^I(s,t;\alpha)=\mathcal{M}_s^I(s,t;\alpha)+\mathcal{M}_t^I(s,t;\alpha)+\mathcal{M}_u^I(s,t;\alpha)+\mathcal{M}_{\mathrm{con}}^I(s,t;\alpha)\;.
\end{equation}
Here $\mathcal{M}_s^I(s,t;\alpha)$, $\mathcal{M}_t^I(s,t;\alpha)$, $\mathcal{M}_u^I(s,t;\alpha)$ are respectively the sum of Mellin amplitudes of all the exchange Witten diagrams in the s-channel, t-channel and u-channel, and  $\mathcal{M}_{\mathrm{con}}^I(s,t;\alpha)$ corresponds to the contact diagrams. More precisely, we include only the singular parts in $\mathcal{M}_s^I(s,t;\alpha)$, $\mathcal{M}_t^I(s,t;\alpha)$, $\mathcal{M}_u^I(s,t;\alpha)$, while the regular polynomial terms will be included in $\mathcal{M}_{\mathrm{con}}^I(s,t;\alpha)$. In the model of gauged supergravity coupled to matter, the number of fields which can appear in an exchange Witten diagram is very small. The exchanged fields are a subset of the spin-$\ell$ fields in the flavor current multiplet and the stress tensor multiplet, subject to the following selection rule:  if $j_R+\ell$ is {\it even}, the exchanged field must be in a {\it symmetric} representation of the flavor symmetry group; if $j_R+\ell$ is {\it odd}, then the exchanged field has to be in an {\it anti-symmetric} representation. Explicitly, in the stress tensor multiplet, where all the fields are flavor symmetry singlets, the following fields are allowed to appear in the exchange Witten diagrams:
 \begin{itemize}
\item a graviton field dual to the stress tensor which has $\Delta=5$, $\ell=2$ and is an R-symmetry singlet;
\item a vector field dual to the R-symmetry current, which has $\Delta=4$, $\ell=1$ and is in $\mathbf{3}$ of $SU(2)_R$;
\item and a scalar field which has $\Delta=3$ and is an R-symmetry singlet.
\end{itemize}
In the flavor current multiplet, where all fields are in the adjoint representation of the flavor symmetry group, we have
 \begin{itemize}
 \item a vector field dual to the flavor symmetry current, which has $\Delta=4$ and is an R-symmetry singlet;
  \item a scalar field dual to the moment map operator, which has $\Delta=3$ and transforms as $\mathbf{3}$ under $SU(2)_R$.
 \end{itemize}
 Notice the dimension-4 $SU(2)_R$ singlet scalar in the flavor current multiplet cannot appear in the exchange diagram because $j_R+\ell$ is odd. 
 
  The R-symmetry irreducible representation of each exchanged field is carried by a multiplicative R-symmetry polynomial 
\begin{equation}
P_{j_R}(2\alpha-1)
\end{equation}
where $P_n(x)$ stands for the Legendre polynomial. These R-symmetry polynomials are the eigenfunctions of the $SU(2)_R$ Casimir. 

For the exchange amplitude, we therefore make the following ansatz\footnote{\label{commentansatz} We include the tree-level exchange of both multiplets  in the ansatz, but it is more general than necessary if  only correlators of the $E_{N_f+1}$ flavor current multiplet are considered. The cubic coupling to the $E_{N_f+1}$ flavor current is of order $N^{-3/2}$ while the coupling to the stress tensor is of order $N^{-5/2}$ \cite{Chang:2017mxc}. The difference of order in $1/N$ means that these two contributions can be separately considered. On the other hand, if external operators are taken from the $SU(2)_M$ (which is part of the isometry of $HS^4$ when the theory is lifted to ten dimensions) mesonic flavor current multiplet , the coupling to the mesonic flavor current multiplet and the stress tensor multiplet are both of the same order $N^{-5/2}$ \cite{Chang:2017mxc}.  Notice also that in any case loop corrections will necessarily be of higher order and wouldn't mix with the tree contributions.}
\begin{equation}
\mathcal{M}_s^I(s,t;\alpha)=\lambda_{S,g} \mathcal{M}_g(s,t)+\lambda_{S,v}P_1(2\alpha-1) \mathcal{M}_v(s,t)+\lambda_{S,s} \mathcal{M}_s(s,t)   \;,\quad {\rm if}\;\; I=\mathbf{1}\;,
\end{equation}
and
\begin{equation}
\mathcal{M}_s^I(s,t;\alpha)= \lambda_{F,v} \mathcal{M}_v(s,t)+  \lambda_{F,s}P_1(2\alpha-1) \mathcal{M}_s(s,t)  \;,\quad {\rm if}\;\; I=\mathbf{adj}\;.
\end{equation}
The s-channel exchange amplitude $\mathcal{M}_s^I(s,t;\alpha)$ is zero otherwise. The parameters $\lambda_{S,g}$,  $\lambda_{S,v}$,  $\lambda_{S,s}$,  $\lambda_{F,v}$ and  $\lambda_{F,s}$ are left as unknowns at this stage.

It is further known that the singular part of exchange Mellin amplitude of a field is the same as that of a conformal block with the same quantum numbers. Using the expressions from \cite{Fitzpatrick:2011hu}, the singular parts of these exchange Mellin amplitudes are
\begin{equation}\label{exchangemellins}\small
\begin{split}
\mathcal{M}_g(s,t)=&\sum_{n=0}^{\infty}\frac{-15\sqrt{\pi}\cos[n\pi]\Gamma[-\frac{5}{2}-n]}{32n!\Gamma^2[\frac{3}{2}-n]}\frac{12 n^2+16 n (t+u-9)+8 (t-9) t+8 (u-9) u+333}{s-(3+2n)}\;,\\
\mathcal{M}_v(s,t)=&\sum_{n=0}^{\infty}\frac{3\sqrt{\pi}\cos[n\pi]\Gamma[-\frac{3}{2}-n]}{16n!\Gamma^2[\frac{3}{2}-n]}\frac{t-u}{s-(3+2n)}\;,\\
\mathcal{M}_s(s,t)=&\sum_{n=0}^{\infty}\frac{-\sqrt{\pi}\cos[n\pi]\Gamma[-\frac{1}{2}-n]}{8n!\Gamma^2[\frac{3}{2}-n]}\frac{1}{s-(3+2n)}
\end{split}
\end{equation}
where $u=12-s-t$\;. 

The t-channel and u-channel Mellin amplitudes $\mathcal{M}_t^I(s,t;\alpha)$, $\mathcal{M}_u^I(s,t;\alpha)$ can be obtained from the s-channel amplitude $\mathcal{M}_s^I(s,t;\alpha)$ using crossing relations\footnote{These relations can be derived as follows. Swapping 1 and 3 interchanges $s$ and $t$ and $\alpha\to\frac{\alpha}{\alpha-1}$. Suppose an exchange amplitude in the s-channel is $\mathcal{M}(s,t;\alpha)$, then the t-channel amplitude is given by $(\alpha-1)^2\mathcal{M}\left(t,s;\frac{\alpha}{\alpha-1}\right)$. The additional  factor $(\alpha-1)^2$ comes from the factor $(t_{12}t_{34})^2$ which we have taken out. However, the flavor irreducible representation $\mathcal{R}_J$ carried by the exchanged field in the s-channel ($ab\to cd$) now becomes the representation $\mathcal{R}_J$ in the t-channel ($cb\to ad$). To project it back to the s-channel 
\begin{equation}
\sum_{a,b,c,d}\frac{P_I^{ab|cd}}{{\rm dim}(\mathcal{R}_I)}\times\left[(\alpha-1)^2P_J^{cb|ad}\mathcal{M}_J\left(t,s;\frac{\alpha}{\alpha-1}\right)\right]\;,
\end{equation}
we acquire the $F_t$\;. The derivation for the u-channel is the same.}
\begin{eqnarray}
\mathcal{M}_t^I(s,t;\alpha)&=&(\alpha-1)^2\sum_J(F_t)_I{}^J\mathcal{M}_s^J\left(t,s;\frac{\alpha}{\alpha-1}\right)\;,\\
\mathcal{M}_u^I(s,t;\alpha)&=&\alpha^2\sum_J(F_u)_I{}^J\mathcal{M}_s^J\left(u,t;\frac{1}{\alpha}\right)\;.
\end{eqnarray}
 With these relations, one can check that the combination
 \begin{equation}
 \mathcal{M}_s^I(s,t;\alpha)+\mathcal{M}_t^I(s,t;\alpha)+\mathcal{M}_u^I(s,t;\alpha)
 \end{equation}
transforms in the desired way (\ref{mellincrossing}) under crossing. This implies that the regular part $\mathcal{M}_{\rm con}^I(s,t;\alpha)$ needs to satisfy the crossing relation (\ref{mellincrossing}) by itself
\begin{equation}\label{mellincrossingcon}
\nonumber \mathcal{M}_{I,{\rm con}}(s,t;\alpha)=(\alpha-1)^2\sum_J (F_t)_I{}^J\mathcal{M}_{J,{\rm con}}\left(t,s;\frac{\alpha}{\alpha-1}\right)= \alpha^2\sum_J (F_u)_I{}^J\mathcal{M}_{J,{\rm con}}\left(u,t;\frac{1}{\alpha}\right)\;.
\end{equation}
We have the following ansatz for $\mathcal{M}_{\rm con}^I(s,t;\alpha)$
\begin{equation}\label{Mcon}
\mathcal{M}_{\rm con}^I(s,t;\alpha)=\sum_{0\leq i\leq 2}\sum_{{\fontsize{4}{3}\selectfont\begin{split}{}&0\leq a,b\leq 1,\\{}&0\leq a+b\leq 1\end{split}}}c^I_{i;a,b}\alpha^i s^at^b\;.
\end{equation}
Note that  $\mathcal{M}_{\rm con}^I(s,t;\alpha)$ is linear in $s$ and $t$ is to ensure that the Mellin amplitude should have a good flat space limit where the theory has at most two derivatives (see \cite{Rastelli:2017udc} for a more detailed discussion).
\subsubsection{Solving the Mellin Superconformal Ward Identities}\label{secsolward}
The Mellin amplitude ansatz we formulated in the last subsection is subject to the Mellin space superconformal Ward identities, whose number depends on the number of flavor irreducible representations appearing in the tensor product of two adjoint representations.. The derivation of such identities has been given in Section \ref{Mellinrepandscfwi}. As was in the previous subsection, we also describe the method of solving superconformal Ward identities in the example of Seiberg theories at large $N$. However the strategy for solving the Mellin superconformal Ward identities applies generally to any ansatz with infinitely many poles, {\it e.g.}, 3d $\mathcal{N}=4$ theories. For ansatz with finitely many poles, such as the case of E-string theory that we will discuss in Section \ref{Estring}, the identities can be easily solved even without using the method outlined here.

Schematically, we will solve these identities in two steps. In the first step we determine the relation among $\lambda_{S,g}$, $\lambda_{S,v}$, $\lambda_{S,s}$ and the relation between $\lambda_{F,v}$ and $\lambda_{F,s}$ -- these parameters appear in the singular part of the Mellin amplitude ansatz. The end result is that we have two independent parameters, $\lambda_S$ and $\lambda_F$, one for each superconformal multiplet. This essentially determines the superconformal block of the stress tensor multiplet and the flavor current multiplet. In the second step, we input the previous solution and solve all the parameters $c^I_{i;a,b}$ in the regular part of the Mellin amplitude ansatz in terms of $\lambda_S$ and $\lambda_F$. Let us now provide more details for this procedure.

For starters, we notice all the poles in the singular part of the Mellin amplitude, {\it i.e.}, $\mathcal{M}_s^I$, $\mathcal{M}_t^I$, $\mathcal{M}_u^I$ are at {\it odd} integers. In the Mellin space superconformal Ward identities, the operator actions (\ref{monomialops}) can shift the position of these poles, but only by an {\it even} integer amount. Another source of poles is the Pochhammer symbols in (\ref{monomialops}), which introduces poles in $u$ at {\it even} integers -- more precisely, at $u=0,2,4$. Because of the opposite parity, we can partially solve the superconformal Ward identities by first focusing on the odd poles and discard the regular piece $\mathcal{M}_{\rm con}^I$ for the time being (which does not contribute to the odd poles).\footnote{In fact, the odd poles in $s$, $t$ and $u$ cannot interfere with each other in solving the Mellin superconformal Ward identities. Therefore we can solve $\lambda_{S,g}$, $\lambda_{S,v}$, $\lambda_{S,s}$, $\lambda_{F,v}$ and  $\lambda_{F,s}$ using only $\mathcal{M}_s^I(s,t)$.} Requiring the odd poles to vanish leads to the following solutions
\begin{eqnarray}
&&\lambda_{S,s}=\lambda_S\;,\quad\lambda_{S,v}=-\frac{4}{3}\lambda_S\;,\quad\lambda_{S,g}=\frac{1}{15}\lambda_S\;,\\
&&\lambda_{F,s}=\lambda_F\;,\quad\lambda_{F,v}=-\frac{1}{3}\lambda_F\;,
\end{eqnarray}
which organizes the exchange Mellin amplitudes into {\it super} exchange Mellin amplitudes for the stress tensor multiplet and the flavor current multiplet respectively. This is the Mellin space version of the method used in \cite{Dolan:2001tt} where superconformal blocks are solved from a linear ansatz of bosonic blocks using superconformal Ward identity.\footnote{See also \cite{Bobev:2017jhk} for an alternative strategy to obtain the superconformal blocks from the superconformal Casimir operator.}

We now plug these solutions into the ansatz (\ref{ansatz}) and further require that the residues of the even integer poles should vanish. The only even integer poles are the aforementioned poles at $u=0$, $u=2$ and $u=4$, with a degree up to 2. Such poles can come from the singular part of the ansatz as well as the regular part via the Pochhammer symbols -- therefore connecting $\lambda_S$, $\lambda_F$ and the parameters $c^I_{i;a,b}$. The vanishing of these residues can only be seen after we resum in $n$ which labels the simple poles of the exchange Mellin amplitudes (\ref{exchangemellins}). These conditions turn out to give enough equations to completely fix all the $c^I_{i;a,b}$ in terms of the two parameters $\lambda_S$, $\lambda_F$. The remaining two coefficients cannot be fixed by symmetries alone, but they have clear physical meanings. In terms of the OPE of two flavor current multiplets, $\lambda_S$ is proportional to the squared OPE coefficient of the stress tensor multiplet; $\lambda_F$ is proportional to that of the flavor current multiplet. Holographically, they are proportional to the gravitational coupling and the gauge coupling. Their values can be determined either directly from supergravity or extracted from deformations of the supersymmetric five-sphere partition function \cite{Chang:2017mxc}.

\subsection{Results for Selected Flavor Groups}\label{selectedG}
In this section we implement the above strategy to compute four-point functions of matter coupled Romans supergravity theories for a number of different flavor symmetry groups: $G= E_1\cong SU(2), E_6, E_7, E_8$. The flavor crossing matrices have been computed in, {\it e.g.}, \cite{Chang:2017xmr,Chang:2017cdx}.

\subsubsection{Flavor Group $E_1\cong SU(2)$}
We start with the simplest example where the flavor group is $E_1\cong SU(2)$. The adjoint representation is $\mathbf{3}$ and in the tensor product we have three representations
\begin{equation}
\mathbf{3}\otimes \mathbf{3}=\underbrace{\mathbf{1}\oplus \mathbf{5}}_{S}\oplus \underbrace{\mathbf{3}}_{A}\;.
\end{equation}
In the above product, $S$ stands for symmetric representations and $A$ stands for antisymmetric representations. Using $(\mathbf{1},\mathbf{5},\mathbf{3})$ as a basis, the flavor crossing matrices are
\begin{equation}
F_t=\left(
\begin{array}{ccc}
 \frac{1}{3} & \frac{5}{3} & 1 \\
 \frac{1}{3} & \frac{1}{6} & -\frac{1}{2} \\
 \frac{1}{3} & -\frac{5}{6} & \frac{1}{2} \\
\end{array}
\right)\;,
\quad F_u=\left(
\begin{array}{ccc}
 \frac{1}{3} & \frac{5}{3} & -1 \\
 \frac{1}{3} & \frac{1}{6} & \frac{1}{2} \\
 -\frac{1}{3} & \frac{5}{6} & \frac{1}{2} \\
\end{array}
\right)\;.
\end{equation}
We then solve the Mellin superconformal Ward identities by following the strategy described in the previous subsection. We find all $c^I_{i;ab}$ parameters in $\mathcal{M}^I_{\rm con}(s,t;\alpha)$ are fixed in terms of $\lambda_S$ and $\lambda_F$
\begin{eqnarray}
\nonumber\mathcal{M}^{\mathbf{1}}_{\rm con}&=&\frac{5\pi\lambda_S(-180+21s+35t)}{576}-\alpha\frac{5\pi\lambda_S(-276+28s+35t)}{288}+\alpha^2\frac{5\pi\lambda_S(-44+7s)}{192}\\
&&+\frac{15\pi}{128}(2\alpha^2-2\alpha-1)\lambda_F\;,\\
\nonumber\mathcal{M}^{\mathbf{5}}_{\rm con}&=&-\frac{5\pi\lambda_S(-18+7t)}{576}+\alpha\frac{5\pi\lambda_S(-102+14s+7t)}{288}-\alpha^2\frac{5\pi\lambda_S(-40+7s)}{192}\\
&&-\frac{15\pi}{256}(2\alpha^2-2\alpha-1)\lambda_F\;,\\
\nonumber\mathcal{M}^{\mathbf{3}}_{\rm con}&=&\frac{5\pi\lambda_S(-150+28s+21t)}{576}-\alpha\frac{5\pi\lambda_S(18+14s-7t)}{288}-\alpha^2\frac{35\pi\lambda_S(-12+s+2t)}{576}\\
&&+\frac{45\pi}{256}(2\alpha-1)\lambda_F\;.
\end{eqnarray}
The fact that we have two independent family of solutions is not surprising. As commented in footnote \ref{commentansatz}, $\lambda_S$ and $\lambda_F$, which are correspondingly proportional to the coupling with the flavor current multiplet and stress tensor multiplet, are of different orders in $1/N$. We should therefore expect that an ansatz exchanging only either multiplet will lead to a non-trivial solution. To fix the last two parameters, we use the result of \cite{Chang:2017mxc}. We find that
\begin{equation}
\lambda_S=-\frac{40}{\pi^2N^{\frac{5}{2}}}\;,\quad \lambda_F=\frac{64}{\pi^2N^{\frac{3	}{2}}}\;.
\end{equation}
More generally, for $0\leq N_f<8$, the values of $\lambda_S$ and $\lambda_F$ are
\begin{equation}
\lambda_S=-\frac{10\sqrt{2}\,{\rm dim}(G)\sqrt{8-N_f}}{3\pi^2N^{\frac{5}{2}}}\;,\quad \lambda_F=\frac{8\sqrt{2}\,h^\vee\sqrt{8-N_f}}{\pi^2N^{\frac{3}{2}}}
\end{equation}
where $h^\vee$ is the dual Coxeter number of the flavor group $G$.

The above computation also admits another interpretation as the four-point correlator of the mesonic current multiplet. Thinking intrinsically in $AdS_6$, the bulk vector multiplet corresponding to the mesonic symmetry {\it a priori} is not different from the $E_{N_f+1}$ symmetries. The $SU(2)_M$ mesonic current multiplet is a singlet under $E_{N_f+1}$. Therefore in the exchange Witten diagrams with four such external operators only the mesonic current multiplet and the stress tensor multiplet will appear. Interpreting the $SU(2)$ in the previous computation as the mesonic symmetry, we see that the relative ratio of $\lambda_S$ and $\lambda_F$ is not fixed by superconformal symmetry alone. However from the 10d perspective, the mesonic symmetry $SU(2)_M$ and the R-symmetry $SU(2)_R$ together make up the isometry $SO(4)=SU(2)_M\times SU(2)_R$ of internal manifold in $AdS_6\times_w HS^4$. Since the R-symmetry current resides in the same multiplet as the stress tensor, we expect that the coupling of the mesonic current multiplet to the stress tensor multiplet and the self-coupling are of the same order -- predicting $\lambda_S/\lambda_F$ is of order one \cite{Chang:2017mxc}.\footnote{$\lambda_F=\frac{320}{3\pi^2}{N^{-\frac{5}{2}}}$ and $\lambda_S/\lambda_F=-\frac{3}{8}$.} This relation perhaps can be seen from computing the $\langle MMtt \rangle$ correlator where $M$ is the moment map operator of the mesonic flavor multiplet and $t$ is the dimension-3 scalar in the stress tensor multiplet. Such a computation can be similarly performed once we have derived the superconformal Ward identity for such four-point functions.

\subsubsection{Flavor Group $E_6$}
Let us move on to the case of $N_f=5$ for which the theory has flavor group $E_6$. The adjoint representation is $\mathbf{78}$ and the tensor product of two adjoint representations gives five flavor channels
\begin{equation}
\mathbf{78}\otimes \mathbf{78}=\underbrace{\mathbf{1}\oplus \mathbf{650}\oplus\mathbf{2430}}_{S}\oplus \underbrace{\mathbf{78}\oplus\mathbf{2925}}_{A}\;.
\end{equation}
We choose the basis vector to be $(\mathbf{1},\mathbf{650},\mathbf{2430},\mathbf{78},\mathbf{2925})$, then we have the following crossing matrices
\begin{equation}
F_t=\left(
\begin{array}{ccccc}
 \frac{1}{78} & \frac{25}{3} & \frac{405}{13} & 1 & \frac{75}{2} \\
 \frac{1}{78} & -\frac{7}{24} & \frac{81}{104} & \frac{1}{4} & -\frac{3}{4} \\
 \frac{1}{78} & \frac{5}{24} & \frac{29}{104} & -\frac{1}{12} & -\frac{5}{12} \\
 \frac{1}{78} & \frac{25}{12} & -\frac{135}{52} & \frac{1}{2} & 0 \\
 \frac{1}{78} & -\frac{1}{6} & -\frac{9}{26} & 0 & \frac{1}{2} \\
\end{array}
\right)\;,\quad 
F_u=\left(
\begin{array}{ccccc}
 \frac{1}{78} & \frac{25}{3} & \frac{405}{13} & -1 & -\frac{75}{2} \\
 \frac{1}{78} & -\frac{7}{24} & \frac{81}{104} & -\frac{1}{4} & +\frac{3}{4} \\
 \frac{1}{78} & \frac{5}{24} & \frac{29}{104} & +\frac{1}{12} & +\frac{5}{12} \\
 -\frac{1}{78} & -\frac{25}{12} & +\frac{135}{52} & \frac{1}{2} & 0 \\
 -\frac{1}{78} & +\frac{1}{6} & +\frac{9}{26} & 0 & \frac{1}{2} \\
\end{array}
\right)\;.
\end{equation}
All the coefficients in the ansatz for $\mathcal{M}^I_{\rm con}$ can be fixed in terms of $\lambda_S$ and $\lambda_F$ by solving the Mellin superconformal Ward identities. The solution is
\begin{eqnarray}
\nonumber\mathcal{M}^{\mathbf{1}}_{\rm con}&=&\frac{5\pi\lambda_S(-5130+546s+1085t)}{14976}-\alpha\frac{5\pi\lambda_S(-9726+1078s+1085t)}{7488}\\
&&+\alpha^2\frac{5\pi\lambda_S(-2144+357s)}{4992}+\frac{15\pi}{128}(2\alpha^2-2\alpha-1)\lambda_F\;,\\
\nonumber\mathcal{M}^{\mathbf{650}}_{\rm con}&=&-\frac{5\pi\lambda_S(-18+7t)}{14976}+\alpha\frac{5\pi\lambda_S(-102+14s+7t)}{7488}-\alpha^2\frac{5\pi\lambda_S(-40+7s)}{4992}\\
&&+\frac{15\pi}{512}(2\alpha^2-2\alpha-1)\lambda_F\;,\\
\nonumber\mathcal{M}^{\mathbf{2430}}_{\rm con}&=&-\frac{5\pi\lambda_S(-18+7t)}{14976}+\alpha\frac{5\pi\lambda_S(-102+14s+7t)}{7488}-\alpha^2\frac{5\pi\lambda_S(-40+7s)}{4992}\\
&&-\frac{5\pi}{512}(2\alpha^2-2\alpha-1)\lambda_F\;,\\
\nonumber\mathcal{M}^{\mathbf{78}}_{\rm con}&=&\frac{5\pi\lambda_S(-150+28s+21t)}{14976}-\alpha\frac{5\pi\lambda_S(18+14s-7t)}{7488}-\alpha^2\frac{35\pi\lambda_S(-12+s+2t)}{14976}\\
&&+\frac{45\pi}{256}(2\alpha-1)\lambda_F\;,\\
\nonumber\mathcal{M}^{\mathbf{2925}}_{\rm con}&=&\frac{5\pi\lambda_S(-150+28s+21t)}{14976}-\alpha\frac{5\pi\lambda_S(18+14s-7t)}{7488}\\
&&-\alpha^2\frac{35\pi\lambda_S(-12+s+2t)}{14976}\;.
\end{eqnarray}

\subsubsection{Flavor Group $E_7$}
The example of $E_7$ is similar to $E_6$. The adjoint representation of $E_7$ is $\mathbf{133}$, and the tensor product of two adjoint representations also contains five irreducible representations
\begin{equation}
\mathbf{133}\otimes \mathbf{133}=\underbrace{\mathbf{1}\oplus \mathbf{1539}\oplus\mathbf{7371}}_{S}\oplus \underbrace{\mathbf{133}\oplus\mathbf{8645}}_{A}\;.
\end{equation}
Choosing the basis vector to be $(\mathbf{1},\mathbf{1539},\mathbf{7371},\mathbf{133},\mathbf{8645})$, we have the following crossing matrices
\begin{equation}
F_t=\left(
\begin{array}{ccccc}
 \frac{1}{133} & \frac{81}{7} & \frac{1053}{19} & 1 & 65 \\
 \frac{1}{133} & -\frac{23}{70} & \frac{78}{95} & \frac{2}{9} & -\frac{13}{18} \\
 \frac{1}{133} & \frac{6}{35} & \frac{61}{190} & -\frac{1}{18} & -\frac{4}{9} \\
 \frac{1}{133} & \frac{18}{7} & -\frac{117}{38} & \frac{1}{2} & 0 \\
 \frac{1}{133} & -\frac{9}{70} & -\frac{36}{95} & 0 & \frac{1}{2} \\
\end{array}
\right)\;,\quad
F_u=\left(
\begin{array}{ccccc}
 \frac{1}{133} & \frac{81}{7} & \frac{1053}{19} & -1 & -65 \\
 \frac{1}{133} & -\frac{23}{70} & \frac{78}{95} & -\frac{2}{9} & +\frac{13}{18} \\
 \frac{1}{133} & \frac{6}{35} & \frac{61}{190} & +\frac{1}{18} & +\frac{4}{9} \\
 -\frac{1}{133} & -\frac{18}{7} & +\frac{117}{38} & \frac{1}{2} & 0 \\
 -\frac{1}{133} & +\frac{9}{70} & +\frac{36}{95} & 0 & \frac{1}{2} \\
\end{array}
\right)\;.
\end{equation}
The superconformal Ward identities fixes all the coefficients in the regular part of the ansatz 
\begin{eqnarray}
\nonumber\mathcal{M}^{\mathbf{1}}_{\rm con}&=&\frac{5\pi\lambda_S(-8760+931s+1855t)}{25536}-\alpha\frac{5\pi\lambda_S(-16656+1848s+1855t)}{12768}\\
&&+\alpha^2\frac{5\pi\lambda_S(-11052+1841s)}{25536}+\frac{15\pi}{128}(2\alpha^2-2\alpha-1)\lambda_F\;,\\
\nonumber\mathcal{M}^{\mathbf{1539}}_{\rm con}&=&-\frac{5\pi\lambda_S(-18+7t)}{25536}+\alpha\frac{5\pi\lambda_S(-102+14s+7t)}{12768}-\alpha^2\frac{5\pi\lambda_S(-40+7s)}{8512}\\
&&+\frac{5\pi}{192}(2\alpha^2-2\alpha-1)\lambda_F\;,\\
\nonumber\mathcal{M}^{\mathbf{7371}}_{\rm con}&=&-\frac{5\pi\lambda_S(-18+7t)}{25536}+\alpha\frac{5\pi\lambda_S(-102+14s+7t)}{12768}-\alpha^2\frac{5\pi\lambda_S(-40+7s)}{8512}\\
&&-\frac{5\pi}{768}(2\alpha^2-2\alpha-1)\lambda_F\;,\\
\nonumber\mathcal{M}^{\mathbf{133}}_{\rm con}&=&\frac{5\pi\lambda_S(-150+28s+21t)}{25536}-\alpha\frac{5\pi\lambda_S(18+14s-7t)}{12768}-\alpha^2\frac{35\pi\lambda_S(-12+s+2t)}{3648}\\
&&+\frac{45\pi}{256}(2\alpha-1)\lambda_F\;,\\
\nonumber\mathcal{M}^{\mathbf{8645}}_{\rm con}&=&\frac{5\pi\lambda_S(-150+28s+21t)}{25536}-\alpha\frac{5\pi\lambda_S(18+14s-7t)}{12768}\\
&&-\alpha^2\frac{5\pi\lambda_S(-12+s+2t)}{3648}\;.
\end{eqnarray}

\subsubsection{Flavor Group $E_8$}\label{5dE8}
Finally, we give the solution for $E_8$. The adjoint representation of $E_8$ is $\mathbf{248}$. In the tensor product of adjoint representations we have
\begin{equation}
\mathbf{248}\otimes \mathbf{248}=\underbrace{\mathbf{1}\oplus \mathbf{3875}\oplus\mathbf{27000}}_{S}\oplus \underbrace{\mathbf{248}\oplus\mathbf{30380}}_{A}\;.
\end{equation}
The representation basis vector is chosen to be $(\mathbf{1},\mathbf{3875},\mathbf{27000},\mathbf{248},\mathbf{30380})$, and the flavor crossing matrices are
\begin{equation}
F_t=\left(
\begin{array}{ccccc}
 \frac{1}{248} & \frac{125}{8} & \frac{3375}{31} & 1 & \frac{245}{2} \\
 \frac{1}{248} & -\frac{3}{8} & \frac{27}{31} & \frac{1}{5} & -\frac{7}{10} \\
 \frac{1}{248} & \frac{1}{8} & \frac{23}{62} & -\frac{1}{30} & -\frac{7}{15} \\
 \frac{1}{248} & \frac{25}{8} & -\frac{225}{62} & \frac{1}{2} & 0 \\
 \frac{1}{248} & -\frac{5}{56} & -\frac{90}{217} & 0 & \frac{1}{2} \\
\end{array}
\right)\;,\quad
F_u=\left(
\begin{array}{ccccc}
 \frac{1}{248} & \frac{125}{8} & \frac{3375}{31} & -1 & -\frac{245}{2} \\
 \frac{1}{248} & -\frac{3}{8} & \frac{27}{31} & -\frac{1}{5} & +\frac{7}{10} \\
 \frac{1}{248} & \frac{1}{8} & \frac{23}{62} & +\frac{1}{30} & +\frac{7}{15} \\
 -\frac{1}{248} & -\frac{25}{8} & +\frac{225}{62} & \frac{1}{2} & 0 \\
 -\frac{1}{248} & +\frac{5}{56} & +\frac{90}{217} & 0 & \frac{1}{2} \\
\end{array}
\right)\;.
\end{equation}
The superconformal Ward identities give the following solution
\begin{eqnarray}
\nonumber\mathcal{M}^{\mathbf{1}}_{\rm con}&=&\frac{5\pi\lambda_S(-16350+1736s+3465t)}{47616}-\alpha\frac{5\pi\lambda_S(-31146+3458s+3465t)}{23808}\\
&&+\alpha^2\frac{5\pi\lambda_S(-20712+3451s)}{47616}+\frac{15\pi}{128}(2\alpha^2-2\alpha-1)\lambda_F\;,\\
\nonumber\mathcal{M}^{\mathbf{3875}}_{\rm con}&=&-\frac{5\pi\lambda_S(-18+7t)}{47616}+\alpha\frac{5\pi\lambda_S(-102+14s+7t)}{23808}-\alpha^2\frac{5\pi\lambda_S(-40+7s)}{15872}\\
&&+\frac{3\pi}{128}(2\alpha^2-2\alpha-1)\lambda_F\;,\\
\nonumber\mathcal{M}^{\mathbf{27000}}_{\rm con}&=&-\frac{5\pi\lambda_S(-18+7t)}{47616}+\alpha\frac{5\pi\lambda_S(-102+14s+7t)}{23808}-\alpha^2\frac{5\pi\lambda_S(-40+7s)}{15872}\\
&&-\frac{\pi}{256}(2\alpha^2-2\alpha-1)\lambda_F\;,\\
\nonumber\mathcal{M}^{\mathbf{248}}_{\rm con}&=&\frac{5\pi\lambda_S(-150+28s+21t)}{47616}-\alpha\frac{5\pi\lambda_S(18+14s-7t)}{23808}-\alpha^2\frac{35\pi\lambda_S(-12+s+2t)}{47616}\\
&&+\frac{45\pi}{256}(2\alpha-1)\lambda_F\;,\\
\nonumber\mathcal{M}^{\mathbf{30380}}_{\rm con}&=&\frac{5\pi\lambda_S(-150+28s+21t)}{47616}-\alpha\frac{5\pi\lambda_S(18+14s-7t)}{23808}\\
&&-\alpha^2\frac{35\pi\lambda_S(-12+s+2t)}{47616}\;.
\end{eqnarray}

\section{Flavor Current Four-Point Function of E-String Theory}\label{Estring}
In this section we apply the Mellin technology to compute flavor current four-point function in E-string theory \cite{Ganor:1996mu,Seiberg:1996vs} which arise in the infrared limit of $N$ M5 branes lying inside an ``end-of-the-world'' M9 brane.  The near horizon limit is $AdS_7\times S^4/\mathbb{Z}_2$ and we will focus on the limit where $N$ is large so that the low energy effective theory can be approximated by supergravity. The $\mathbb{Z}_2$ action breaks the $S^4$ isometry $SO(5)$ into $SO(4)=SU(2)_L\times SU(2)_R$. The $SU(2)_R$ is identified with the R-symmetry group of the superconformal $OSp(6,2|2)$ while the other $SU(2)_L$ becomes a flavor symmetry. The $\mathbb{Z}_2$ acting on $AdS_7\times S^4/\mathbb{Z}_2$ has a fixed point locus $AdS_7\times S^3$ and on this locus we have a 10d $\mathcal{N}=1$ $E_8$ SYM multiplet from the Ho\v{r}ava-Witten compactification \cite{Horava:1996ma,Horava:1995qa}. One can perform KK reduction of theory to $AdS_7$ and further truncates it to the massless multiplets. The truncated theory is $\mathcal{N}=2$ gauged supergravity coupled to matter on $AdS_7$ \cite{Mezincescu:1984ta,Bergshoeff:1985mr}. This theory contains stress tensor multiplet and $E_8$ vector multiplet and will be relevant to our computation of the flavor current multiplet four-point function.

Let us discuss the multiplets in more details. The stress tensor multiplet is neutral under flavor symmetries. The relevant fields are a dimension-6 $SU(2)_R$ singlet graviton field, a dimension-5 vector field in the $\mathbf{3}$ of $SU(2)_R$ and a dimension-4 $SU(2)_R$ singlet scalar. The flavor current multiplet transforms in the adjoint representation of the flavor current. The relevant fields are a dimension-4 scalar  in the $\mathbf{3}$ of $SU(2)_R$ and a dimension-5 vector in the singlet representation of the R-symmetry. The quantum numbers of these multiplets satisfies a special condition, namely, $\tau+2m=2\Delta$ where $\tau$ is the conformal twist of the exchanged field, $\Delta$ is the dimension of the moment map operator and $m$ is a non-negative integer. This conspiracy of quantum numbers leads to a simplification in the exchange Mellin amplitude: the infinite series of simple poles truncates into finitely many, as we explained in \cite{Rastelli:2017udc}. Therefore, using the Mellin expression of conformal blocks \cite{Fitzpatrick:2011hu}, the singular parts of the exchange amplitudes are just rational functions\footnote{Alternatively, these Mellin amplitudes can also be obtained from first evaluating the Witten diagrams in position space using the method of \cite{DHoker:1999aa}. }
\begin{equation}\label{exchangemellins6d}\small
\begin{split}
 \mathcal{M}_g(s,t)=&\frac{5 s^2+10 s (t-10)+2 \left(5 t^2-80 t+344\right)}{2 (s-4)}+\frac{5 s^2+10 s (t-11)+2 \left(5 t^2-80 t+368\right)}{8 (s-6)}\;,\\
 \mathcal{M}_v(s,t)=&\frac{s+2 t-16}{2 (s-4)}+\frac{s+2 t-16}{6 (s-6)}\;,\\
\mathcal{M}_s(s,t)=& \frac{1}{s-4}+\frac{1}{2(s-6)}\;.
\end{split}
\end{equation}
Together with the same $\mathcal{M}^I_{\rm con}$ (\ref{Mcon}) for the regular part, the ansatz for the Mellin amplitude simplifies into a simple rational function.

This rationality makes solving the Mellin superconformal Ward identities particularly straightforward. We find that the exchange parameters $\lambda_{S,g}$, $\lambda_{S,v}$, $\lambda_{S,s}$, $\lambda_{F,v}$ and  $\lambda_{F,s}$ are solved in terms of two parameters $\lambda_S$ and $\lambda_F$
\begin{eqnarray}
&&\lambda_{S,s}=\lambda_S\;,\quad\lambda_{S,v}=-\frac{5}{2}\lambda_S\;,\quad\lambda_{S,g}=\frac{1}{6}\lambda_S\;,\\
&&\lambda_{F,s}=\lambda_F\;,\quad\lambda_{F,v}=-\frac{1}{2}\lambda_F\;\;.
\end{eqnarray}
The contact parameters $c^I_{i;a,b}$ are also solved in terms of  $\lambda_S$ and $\lambda_F$ but the solution depends on the flavor current correlator we consider. In the next two subsections, we record the solution of $c^I_{i;a,b}$ for $E_8$ flavor current and the $SU(2)_L$ flavor current.

\subsection{$E_8$ Flavor Current}
The exchange Witten diagrams of the $E_8$ flavor current multiplet four-point function involve only the $E_8$ flavor current multiplet and the stress tensor multiplet. As in the 5d example in Section \ref{5dE8}, there are five flavor channels and the flavor crossing matrices were given there. Solving the Mellin superconformal Ward identities, we find the following solution for the contact part

\begin{eqnarray}
\nonumber\mathcal{M}^{\mathbf{1}}_{\rm con}&=&\frac{5\lambda_S(-15856+1240s+2475t)}{11904}-\alpha\frac{5\lambda_S(-29664+2470s+2475 t)}{5952}\\
&&+\alpha^2\frac{5\lambda_S(-19728+2465s)}{11904}+\frac{1}{3}(2\alpha^2-2\alpha-1)\lambda_F\;,\\
\nonumber\mathcal{M}^{\mathbf{3875}}_{\rm con}&=&-\frac{5\lambda_S(-16+5t)}{11904}+\alpha\frac{5\lambda_S(-96+10s+5t)}{5952}-\alpha^2\frac{5\lambda_S(-112+15s)}{11904}\\
&&+\frac{1}{15}(2\alpha^2-2\alpha-1)\lambda_F\;,\\
\nonumber\mathcal{M}^{\mathbf{27000}}_{\rm con}&=&-\frac{5\lambda_S(-16+5t)}{11904}+\alpha\frac{5\lambda_S(-96+10s+5t)}{5952}-\alpha^2\frac{5\lambda_S(-112+15s)}{11904}\\
&&-\frac{1}{90}(2\alpha^2-2\alpha-1)\lambda_F\;,\\
\nonumber\mathcal{M}^{\mathbf{248}}_{\rm con}&=&\frac{5\lambda_S(-144+20s+15t)}{11904}-\alpha\frac{5\lambda_S(16+10s-5t)}{5952}-\alpha^2\frac{25\lambda_S(-16+s+2t)}{11904}\\
&&+\frac{1}{2}(2\alpha-1)\lambda_F\;,\\
\nonumber\mathcal{M}^{\mathbf{30380}}_{\rm con}&=&\frac{5\lambda_S(-144+20s+15t)}{11904}-\alpha\frac{5\lambda_S(16+10s-5t)}{5952}\\
&&-\alpha^2\frac{25\lambda_S(-16+s+2t)}{11904}\;.
\end{eqnarray}
The remaining two parameters are determined using the result from \cite{Chang:2017xmr}. We have
\begin{equation}
\lambda_S=-\frac{372}{5N^3}\;,\quad \lambda_F=\frac{30}{N^2}\;.
\end{equation}

\subsection{$SU(2)_L$ Flavor Current}
Let us consider the four-point function of the moment map operator in the $SU(2)_L$ flavor current multiplet. Due to the flavor symmetry selection rule, only the $SU(2)_L$ flavor current multiplet and the stress tensor multiplet can appear in the exchange Witten diagram. The correlator contains three flavor channels and the contact parameters are solved to be

\begin{eqnarray}
\nonumber\mathcal{M}^{\mathbf{1}}_{\rm con}&=&\frac{5\lambda_S(-176+15s+25t)}{144}-\alpha\frac{5\lambda_S(-264+20s+25t)}{72}+\alpha^2\frac{5\lambda_S(-128+15s)}{144}\\
&&+\frac{1}{3}(2\alpha^2-2\alpha-1)\lambda_F\;,\\
\nonumber\mathcal{M}^{\mathbf{5}}_{\rm con}&=&-\frac{5\lambda_S(-16+5t)}{144}+\alpha\frac{5\lambda_S(-96+10s+5t)}{72}-\alpha^2\frac{5\lambda_S(-112+15s)}{144}\\
&&-\frac{1}{6}(2\alpha^2-2\alpha-1)\lambda_F\;,\\
\nonumber\mathcal{M}^{\mathbf{3}}_{\rm con}&=&\frac{5\lambda_S(-144+20s+15t)}{144}-\alpha\frac{5\lambda_S(16+10s-5t)}{72}-\alpha^2\frac{25\lambda_S(-16+s+2t)}{144}\\
&&+\frac{1}{2}(2\alpha-1)\lambda_F\;.
\end{eqnarray}
The remaining two parameters are of the same order because $SU(2)_L$ and $SU(2)_R$ have the same origin from the isometry of the internal manifold. Using the result from \cite{Chang:2017xmr},   we find that
\begin{equation}
\lambda_S=-\frac{372}{5N^3}\;,\quad \lambda_F=\frac{3}{N^3}\;,
\end{equation}

\acknowledgments
X.Z. thanks Wolfger Peelaers, Leonardo Rastelli and Yifan Wang for useful discussions and comments on the manuscript. X.Z. also thanks the Leinweber Center for Theoretical Physics of the University of Michigan for hospitality where the manuscript was partly prepared. The work of X.Z. is supported in part by NSF Grant PHY-1620628.

\bibliography{mellin8supercharges} 
\bibliographystyle{utphys}

\end{document}